# On the nature of time in time-dependent expansionary processes

Laurence Francis Lacey

Lacey Solutions Ltd, Skerries, County Dublin, Ireland

Mon June 01 2021

# On the nature of time in time-dependent expansionary processes

## Abstract

For an expansionary process, the size of the expansion space will increase. If the expansionary process is time-dependent, time (t) will increase as a function of the increase in the size of the expansion space. A statistical information entropy methodology was used to investigate the properties of time-dependent expansionary processes both in theory and through examples. The primary objective of this paper was to investigate whether there is a universal measure of time (T) and how it relates to process related time (t), that is specific to any given time-dependent expansionary process. It was found that for such time-dependent processes, time (t) can be rescaled to time (T) such that, T and the information entropy (H(T)) of the expansionary process are the same, and directly related to the increase in the size of the expansion space.

# 1. Introduction

An expansionary process is one that increases in size. A time-dependent expansionary process is one that increases in size with time. A statistical information entropy methodology was used to investigate the properties of time-dependent expansionary processes.

Information entropy (H) will be used in the following form:

$$H = -\sum_{i}^{n} p_i \log_e p_i$$

H = 0 for a random variable (X) with a determined outcome (i.e., X = $x_0$), p($x_0$) = 1 [1].

The primary objective of this paper is to investigate the properties of time in time-dependent expansionary processes. In particular, this paper will investigate whether there is a universal measure of time (T) and how it relates to process related time (t), that is specific to any given time-dependent expansionary process.

# 2. Methods

## 2.1 Expansion process

There is a random variable (X) with a determined outcome (i.e., X = $x_0$), p($x_0$) = 1. Consider $x_0$ to have a discrete uniform distribution over the integer interval [1, s], where the size of the sample space (s) = 1, in the initial state, such that p($x_0$) = 1, H=0. It can be shown (Appendix) that:

$$s(n) = 2^n = \exp(\log_e 2 \times n)$$

where, n is the number of expanding sample space doublings. The corresponding probability and information entropy of the system, as a function of n, is given by:

$$p(x_0|s_n) = \exp(-log_e 2 \text{ x n})$$

$$H(n) = n \text{ x } log_e 2$$

## 2.2 Time-dependent expansion processes

Three different time-dependent expansion processes will be described:

**(1) Exponential time-dependent expansion:** If the expansion occurs at an exponential rate (rate constant = λ) with time (t), such that:

$$t = n \text{ x } \frac{log_e(2)}{\lambda}, \text{giving } n = \frac{\lambda \text{ x t}}{log_e(2)}$$

By substituting t for n, gives:

$$s(t) = \exp(\lambda \text{ x t})$$

$$p(x_0|t) = \exp(-\lambda \text{ x t})$$

$$H(t) = \lambda \text{ x t}$$

**(2) Power time-dependent expansion:** If the expansion occurs at a rate with time (t) such that:

$$t = 2^n, \text{giving } n = \frac{log_e(t)}{log_e(2)}$$

By substituting t for n, gives

$$s(t) = t$$

$$p(x_0|t) = t^{-1}$$

$$H(t) = log_e(t)$$

This is the characterization of a power function expansion and this functional form of expansion can be show for any expansion in which:

$$t^a = 2^n, \text{giving } t = 2^{n^{1/a}} \text{ and } n = \frac{a \times log_e(t)}{log_e(2)} \text{ for a > 0. for a > 0.}$$

In this case, substituting t for n, gives:

$$s(t) = t^a$$

$$p(x_0|t) = t^{-a}$$

$$H(t) = log_e(t^a)$$

**(3) Double exponential time-dependent expansion:** If the expansion occurs at a rate with time (t) such that:

$$t = \lambda^{-1} \times log_e\left(\frac{n}{a}\right), \text{giving } n = a \times log_e(2) \times exp(\lambda \times t), \text{ for a > 0.}$$

In this case, substituting t for n, gives:

$$s(t) = exp(log_e(2^a) \times exp(\lambda \times t))$$

$$p(x_0|t) = exp(log_e(2^a) \times exp(-\lambda \times t))$$

$$H(t) = log_e(2^a) \times exp(\lambda \times t)$$

## 2.3 Time rescaling of time-dependent expansion processes

**(1) Exponential time-dependent expansion:** If the expansion occurs at an exponential rate (rate constant = λ) with time (t), such that:

$$t = n \, x \, \frac{\log_e(2)}{\lambda}, \text{giving } n = \frac{\lambda \, x \, t}{\log_e(2)}$$

This gives:

$$T = \lambda \, x \, t = n \, x \, \log_e(2)$$

**(2) Power time-dependent expansion:** If the expansion occurs at a rate with time (t) such that:

$$t^a = 2^n, \text{giving } t = 2^{n^{1/a}} \text{ and } n = \frac{a \, x \, \log_e(t)}{\log_e(2)} \text{ for a > 0.}$$

This gives:

$$T = a \, x \, \log_e(t) = n \, x \, \log_e(2)$$

**(3) Double exponential time-dependent expansion:** If the expansion occurs at a rate with time (t) such that:

$$t = \lambda^{-1} \, x \, \log_e\left(\frac{n}{a}\right), \text{giving } n = a \, x \, \log_e(2) \, x \, \exp(\lambda \, x \, t), \text{ for a > 0.}$$

This gives:

$$T = a \, x \, \log_e(2) \, x \, \exp(\lambda \, x \, t) = n \, x \, \log_e(2)$$

## 3. Results

In situations where the time-dependent expansionary process can be expressed as:

$$T = f(t) = n \times \log_e(2)$$

the expansion process can be quantified in terms of the size of the expanding sample space (s(T)), and the associated probability (p(T)), and information entropy (H(T)) as shown in Figure 1.

Figure 1: Semi-log plot of the expanding sample space (s(T)), and the associated probability (p(T)), and information entropy (H(T)) with T

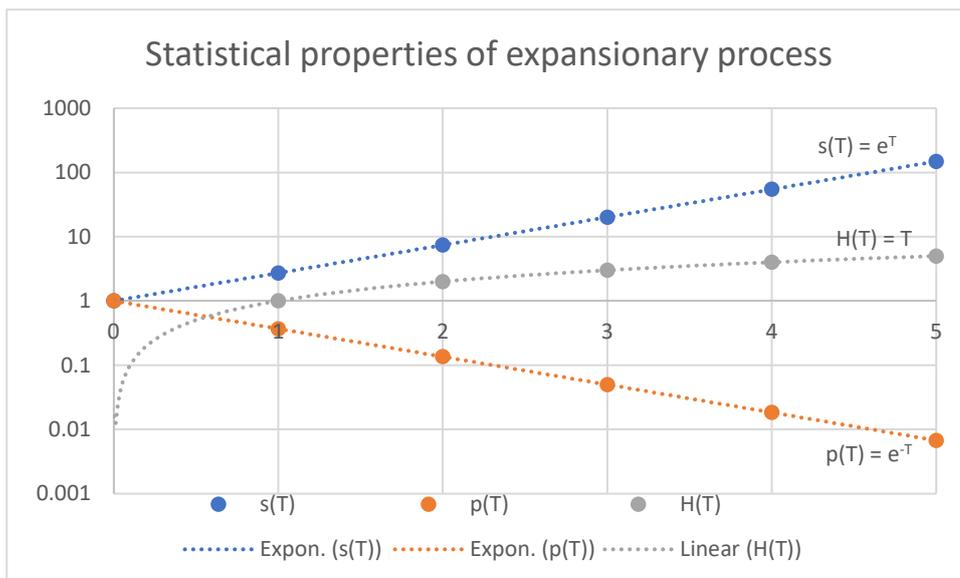

As can be seen, following time rescaling:

$$H(T) = T$$

Three examples of time-dependent expansionary process will be given, involving separate, simultaneous, independent processes.

**Example 1: A time-dependent expansion involving two simultaneous, independent processes, each of different functional form**

The overall time-dependent expansion process P(t) involves two independent processes P1(t), P2(t), with P1(t) being exponential in nature and P2(t) being power in nature.

$$p(x_0|P1, t) = \exp(-t)$$

$$p(x_0|P2, t) = t^{-1}$$

For the overall process P(t):

$$p(x_0|t) = p(x_0|P1, t) \times p(x_0|P2, t) = \exp(-t) \times t^{-1}$$

Setting:

$$T = t + \log_e(t)$$

gives

$$p(x_0|t) = \exp(-T)$$

Please note that for this particular process, $p(x_0|t)\ is\ set = 1$, when t=1, requiring setting t = t-1 for P1. The relationship between process time (t) and T is given in Figure 2.

Figure 2: A plot of t with T for an example of a time-dependent expansion involving two simultaneous, independent processes, each of different functional form

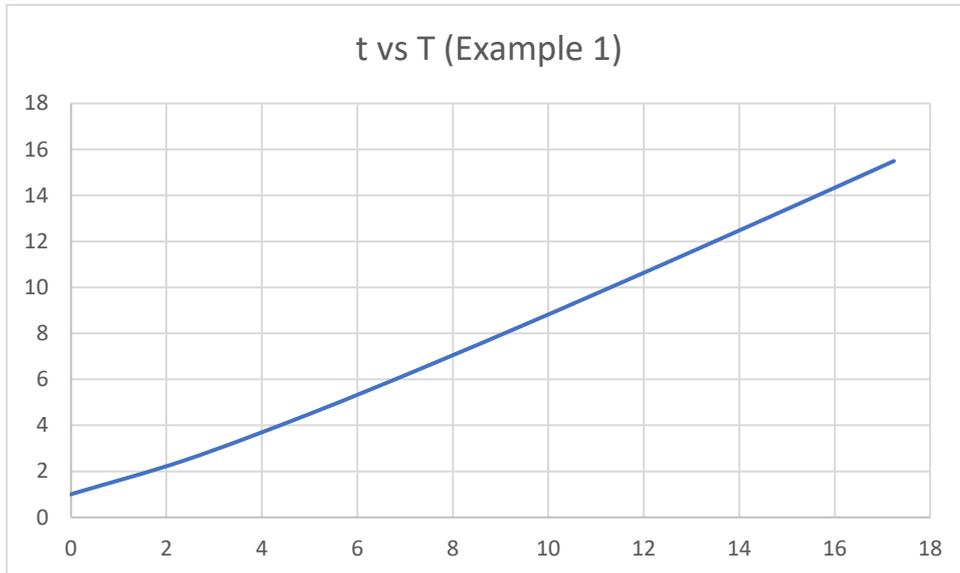

**Example 2: A time-dependent expansion involving three simultaneous, independent processes**

The overall time-dependent expansion process P(t) involves three independent processes P1(t), P2(t), P3(t) with P1(t) being exponential in nature, P2(t) and P3(t) being power in nature.

$$p(x_0|P1, t) = \exp(-t)$$

$$p(x_0|P2, t) = t^{-1/2}$$

$$p(x_0|P3, t) = t^{-2/3}$$

For the overall process P(t):

$$p(x_0|t) = p(x_0|P1, t) \; x \; p(x_0|P2, t) \; x \; p(x_0|P3, t)$$

$$= \exp(-t) \, x \, t^{-1/2} \, x \, t^{-2/3}$$

Setting:

$$T = t + \left(\frac{1}{2}\right) x \, log_e(t) + \left(\frac{2}{3}\right) x \, log_e(t) = t + \left(\frac{7}{6}\right) x \, log_e(t)$$

gives

$$p(x_0|t) = \exp(-T)$$

Please note that for this particular process, $p(x_0|t) \; is \; set = 1$, when t=1, requiring setting t = t-1 for P1. The relationship between process time (t) and T is given in Figure 3.

Figure 3: A plot of t with T for an example of a time-dependent expansion involving three simultaneous, independent processes

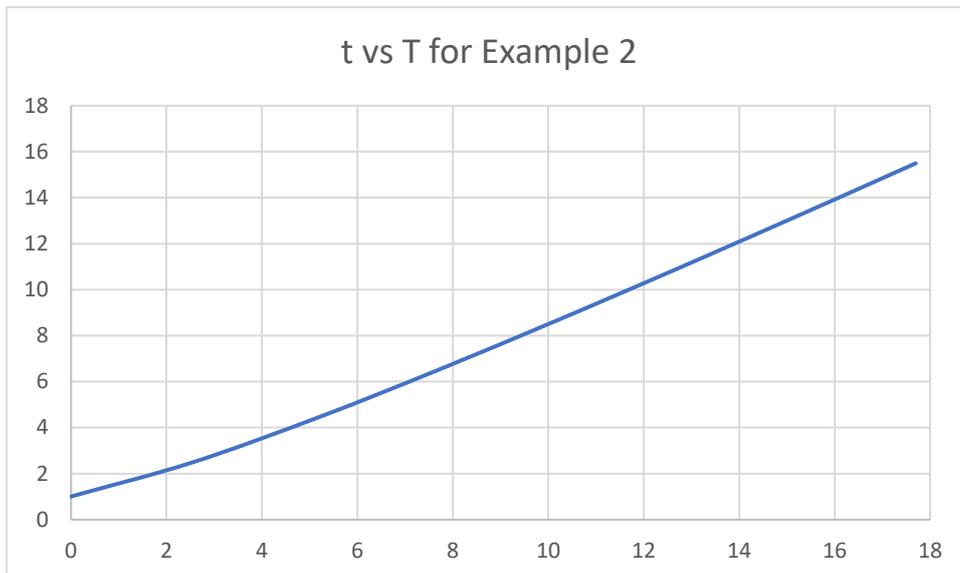

**Example 3: A time-dependent expansion involving three simultaneous, independent processes, each of different functional form**

The overall time-dependent expansion process P(t) involves three independent processes P1(t), P2(t), P3(t) with P1(t) being exponential in nature, P2(t) being power in nature, and P(3) being double exponential in nature.

$$p(x_0|P1, t) = \exp(-t)$$

$$p(x_0|P2, t) = t^{-1}$$

$$p(x_0|P3, t) = \exp(log_e(2) \; x \exp(-t))$$

For the overall process P(t):

$$p(x_0|t) = p(x_0|P1, t) \; x \; p(x_0|P2, t) \; x \; p(x_0|P3, t)$$

$$= \exp(-t) \; x \; t^{-1} \; x \exp(log_e(2) \; x \exp(-t))$$

Setting:

$$T = t + log_e(t) + (log_e(2) \; x \exp(t))$$

gives

$$p(x_0|t) = \exp(-T)$$

Please note that for this particular process, $p(x_0|t) \; is \; set = 1$, when t=1, requiring setting t = t-1 for P1 and t = t-12.5 for P3, respectively. The relationship between process time (t) and T is given in Figure 4.

Figure 4: A plot of t with T for an example of a time-dependent expansion involving three simultaneous, independent processes, each of different functional form

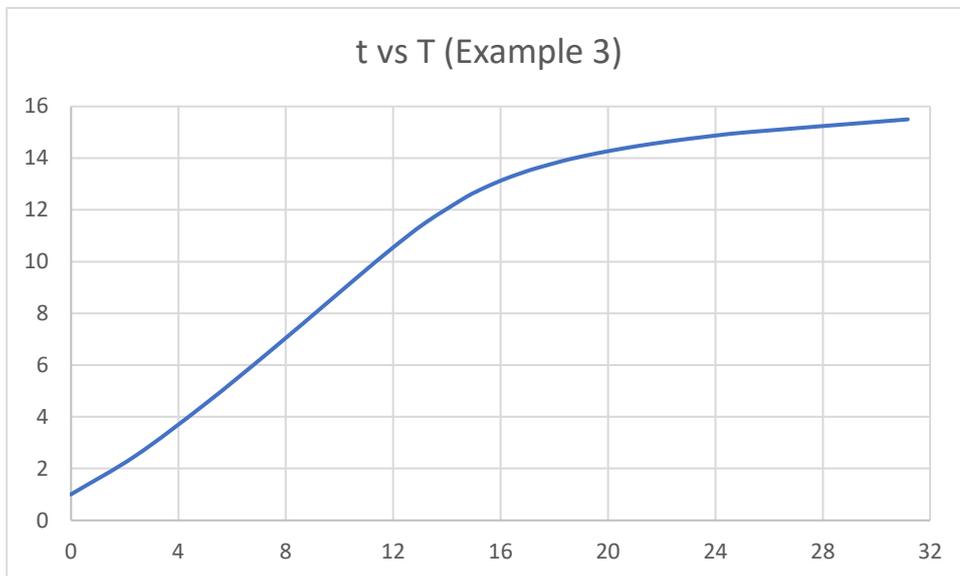

## 4. Discussion

The operational definition of time (e.g., the passage of a free-swinging pendulum) does not address what the fundamental nature of time is. It does not address why events can happen forward and backward in space, whereas events only happen in the forward direction with the progress of time [2]. This "one-way direction" of time is sometimes called the "arrow of time" [3].

For an expansionary process, the size of the expansion space will increase. If the expansionary process is time-dependent, time will increase as a function of the increase in the size of the expansion space.

This paper has identified two related measures of time for expansionary processes:

(1) A "fundamental" measure of time (T) that is directly related to the increase in the size of the expansion space, i.e.,

$$T = H(T) = \log_e(2) \: x \: n$$

where, n is the number of doublings in the size of the expansion space

(2) A process-specific measure of time (t), where,

$$f(t) = T$$

As can be seen from the examples explored, the relationship between t and T can be approximately linear, especially as T becomes large (Examples 1 and 2) or very non-linear (Example 3).

Example 2 may be of relevance to aspects of the Friedmann–Lemaître–Robertson–Walker (FLRW) metric, as the following relationships have been found to apply [5]. However, any such relevance is outside the scope of this paper.

For a radiation-dominated universe, the evolution of the scale factor in the FLRWr metric (a(t)):

$$a(t) \propto t^{1/2}$$

Similarly, for a matter-dominated universe, the evolution of the scale factor is:

$$a(t) \propto t^{2/3}$$

And for a dark energy-dominated universe, the evolution of the scale factor is:

$$a(t) \propto \exp(H_0 \: x \: t)$$

where, $H_o$ is the value of the Hubble "constant"

Whether there can be expansionary processes where the relationship:

$$T = H(T) = f(t) = n \times \log_e(2)$$

does not hold is unknown to the author, but the relatively complex examples explored would seem to confirm the general nature of this relationship.

## 5. Conclusion

For an expansionary process, the size of the expansion space will increase. If the expansionary process is time-dependent, time will increase as a function of the increase in the size of the expansion space. For time-dependent expansionary processes, time (t) can be rescaled to time (T) such that:

$$T = H(T) = f(t) = n \times \log_e(2)$$

where, n is the number of doublings of the expansion space. Following time rescaling, T and the information entropy (H(T)) of the expansionary process are the same.

## Supplementary materials

There is an Appendix giving the statistical characterisation of an expansion process.

## Acknowledgements

The author gratefully acknowledges those academic and non-academic people who kindly provided feedback on earlier drafts of this paper. No financial support was received for any aspect of this research.

# Appendix: Statistical characterisation of an expansion process

There is a random variable (X) with a determined outcome (i.e., X = $x_0$), $p(x_0) = 1$. Consider $x_0$ to have a discrete uniform distribution over the integer interval [1, s], where the size of the sample space (s) = 1, in the initial state, such that $p(x_0) = 1$, H=0. I

For a sample space of size, $s_0 > 1$, let $x_0$ have a discrete uniform distribution over the integer interval [1, $s_0$], giving:

$$p(x_0) = \frac{1}{s_0}$$

where, $p(x_0)$ is the probability mass function over the integer interval [1, $s_0$]. The cumulative probability of $p(x_0)$ over [1, $s_0$] is equal to 1. Let $p(x_0|s_0)$ be the <u>cumulative probability</u> of $p(x_0)$ over [1, $s_0$], i.e.,

$$p(x_0|s_0) = 1$$

Double the sample space in which $x_0$ will be uniformly distributed, i.e., $s_1 = 2 \times s_0$, and partition the new sample space into 2 equally sized, mutually exclusive sample spaces, each of size $s_0$. Now:

$$p(x_0) = \frac{1}{s_1} = \frac{1}{2 \times s_0}$$

The cumulative probability of $x_0$, in any <u>one</u> of the two partitions of size $s_0$, is now:

$$p(x_0|s_1) = \tfrac{1}{2}$$

Repeating the process a second time, the cumulative probability of $x_0$, in any <u>one</u> of the four partitions, each of size $s_0$, is now:

$$p(x_0|s_2) = \tfrac{1}{4}$$

Repeating the process n times, the cumulative probability, for $x_0$ in any <u>given</u> partition of size $s_0$ is:

$$p(x_0|s_n) = (\tfrac{1}{2})^n$$

This is an exponential process, with the sample space expanding at an exponential rate, with:

$$s(n) = 2^n = \exp(\log_e 2 \times n)$$

The corresponding probability and information entropy of the system, as a function of n, is given by:

$$p(x_0|s_n) = \exp(-\log_e 2 \times n)$$

$$H(n) = -2^n \times (\tfrac{1}{2})^n \times \log_e (\tfrac{1}{2})^n = n \times \log_e 2$$

Further methodological details are given in [6].